\documentclass[12pt,preprint]{aastex}
\usepackage{amsmath}
\usepackage{color}
\usepackage{color}

\def\jcap{JCAP}
\def\beq{\begin{equation}}
\def\eeq{\end{equation}}
\def\ben{\begin{eqnarray}}
\def\een{\end{eqnarray}}
\def\munit{M_{\odot}}
\def\ms{M_{\star}}
\def\lam{\lambda}
\def\be{{\bf e}_{1}}
\def\bjd{{\bf j}_{d}}
\def\bjs{{\bf j}_{s}}
\def\cosaa{\langle\cos\alpha\rangle}
\def\costa{\langle\cos\theta\rangle}
\def\cosa{\cos\alpha}
\def\cost{\cos\theta}
\def\tm{M_{\rm tot}}

\begin{document}
\title{How the Galaxy Stellar Spins Acquire a Peculiar Tidal Connection?}
\author{Jounghun Lee\altaffilmark{1}, Jun-Sung Moon\altaffilmark{2},
Suk-Jin Yoon\altaffilmark{2}}
\altaffiltext{1}{Astronomy Program, Department of Physics and Astronomy, 
Seoul National University, Seoul 08826, Republic of Korea 
\email{jounghun@astro.snu.ac.kr}}
\altaffiltext{2}{Department of Astronomy, Yonsei University, Seoul, 03722, Republic of Korea 
\email{moonjs@yonsei.ac.kr,sjyoon0691@yonsei.ac.kr}}

\begin{abstract}
We explore how the galaxy stellar spins acquire a peculiar tendency of being aligned with the major principal axes of the 
local tidal fields, in contrast to  their DM counterparts which  tend to be perpendicular to them, regardless of their masses. 
Analyzing the halo and subhalo catalogs from the IllustrisTNG 300 hydrodynamic simulations at $z\le 1$, we determine the 
cosines of the alignment angles, $\cos\alpha$, between the galaxy stellar and DM spins. Creating four $\cosa$-selected samples 
of the galaxies and then controlling them to share the same density and mass distributions, we determine the average strengths 
of the alignments between the galaxy stellar spins and the tidal tensor major axes over each sample.  
It is clearly shown that at $z\le 0.5$ the more severely the galaxy stellar spin directions deviate from the DM counterparts, 
the stronger the peculiar tidal alignments become. Taking the ensemble averages of such galaxy properties as central blackhole to stellar 
mass ratio, specific star formation rate, formation epoch, stellar-to-total mass ratio, velocity dispersions, average metallicity, and degree of the 
cosmic web anisotropy over each sample,  we also find that all of these properties exhibit either strong correlations or anti-correlations with $\cos\alpha$. 
Our results imply that the peculiar tidal alignments of the galaxy stellar spins may be caused by anisotropic occurrence of some baryonic process 
responsible for discharging stellar materials from the galaxies along the tidal major directions at $z<1$. 
\end{abstract}
\section{Introduction}\label{sec:intro}

The intrinsic spin alignments of dark matter (DM) halos with the principal axes of the local tidal fields have so far been 
extensively studied in numerous literatures for various purposes. The main purpose was to distinguish them from the 
extrinsic spin alignments of galaxies caused by weak gravitational lensing effect and sort them out as systematic errors 
in the cosmic shear analysis \citep[for a comprehensive review, see][]{wl_review15}. 
The other primary purpose was to understand the origin  of galaxy angular momentum which holds one of the key clues to understanding 
the formation and evolution of galaxies \citep[see][for a review]{align_review15}. 

Another purpose that has recently gain considerable attention is to use them as a probe of the initial conditions of the universe. 
For instance, \citet{yu-etal20} claimed that the intrinsic spin alignments of galaxies could be used to examine the 
violation of chiral symmetry of weak interactions in the early universe. 
\citet{TO20} suggested that the two-point correlations of halo intrinsic alignments should be a sensitive tracer of the cosmic 
growth history \citep[see also][]{chu-etal21}. 
\citet{LL20} found that the dark energy (DE) equation of state can in principle be constrained by the mass scales at which the preferential 
alignments of DM halo spins with the tidal principal axes exhibit transitions. 
The power of the intrinsic spin alignments of DM halos as a probe of cosmology is that as a near-field, galactic-scale 
diagnostics, it is capable of distinguishing between the standard $\Lambda$CDM (cosmological constant 
$\Lambda$ and cold DM) and very similar alternative cosmologies, which the conventional diagnostics fail to tell apart 
\citep[e.g.,][]{lee-etal20}. 

To achieve this goal of probing the background cosmology with the intrinsic spin alignments of DM halos in practice, however, 
two prerequisite questions had to be answered. The first one was how well the {\it observable} stellar spins of galaxies are aligned with 
the {\it unobservable} DM spins. 
The second question, a more critical one, was that if the galaxy stellar spins are not perfectly aligned with the DM spins, then do 
they still retain a significant connection to the local tidal fields as their DM spins do? 
The intrinsic spin alignments of DM halos depend on the initial conditions of the universe through their connections to the local tidal fields, 
which plays an intermediary role.  If the deviation of the galaxy stellar angular momentum from those of the DM counterparts accompanies complete 
loss of the intermediary tidal connections, then it would no longer be feasible to probe the background cosmology with the intrinsic spin alignments 
of galaxies. 

The first question has been indisputably answered by multiple numerical experiments based on hydrodynamical simulations, which 
consistently found substantial misalignments between galaxy stellar and DM spins 
\citep[e.g.,][and references therein]{hah-etal10,vel-etal15,ten-etal17,zju-etal17,sou-etal20}.  The discharge of stellar materials out of the galaxies
by the galactic winds has been suggested as the main mechanism of inducing this misalignments \citep[see][and references therein]{ten-etal17}.

Unlike the first one, however, the second question still awaits a conclusive answer, which requires expensive high-resolution {\it cosmological} hydrodynamical 
simulations.  While several works reported that the alignment tendencies of the galaxy stellar spins with the local filaments were found 
similar to those of DM spins at low redshifts $z\le 0.5$ \citep[e.g.,][]{wan-etal18,kra-etal20}, 
several others claimed that the nonlinear baryonic process responsible for the deviation of the galaxy stellar spin directions from those of DM 
counterparts must have completely destroyed any tidal connection, randomizing the galaxy stellar spin vectors relative to the local filaments
at $z\le 0.5$ \citep[e.g.,][]{cod-etal18,ballet2,shi-etal21}.  The difference in the identifications of the filaments has been suggested as a possible 
reason for the discrepant results \cite{kra-etal20}. 

Very recently, a new inkling of the answer to the second question has been found by \citet{lee-etal21} who 
investigated the evolution of the spin transition mass of DM subhalos by analyzing the IllustrisTNG cosmological hydrodynamical 
simulations \citep{illustris19}. According to their results, at low redshifts $z\le 0.5$, the stellar spins of massive galaxies with stellar masses  
$M_{\star}\ge 10^{10}\,M_{\odot}$ exhibit a {\it peculiar} tendency to be aligned with the {\it major} principal axes of the local tidal fields, 
to which those of their DM counterparts are always perpendicular in the entire stellar mass range.  
Yet, at higher redshifts $z>1$, the stellar and DM spins of massive galaxies yield the same tendency 
of being preferentially aligned with the tidal intermediate principal axes and perpendicular to the tidal major axes. 
Unlike the previous studies which focused on the alignments of the galaxy stellar spins only with the elongated axes of filaments
\citep[e.g.,][]{cod-etal18,ballet2,kra-etal20}, \citet{lee-etal21} considered their alignments with all of the three principal directions of the 
tidal fields, which led them to find this lurking difference between the stellar and DM spin alignment tendencies. 
The results of \cite{lee-etal21} implied that at $z\le 0.5$ the baryonic process responsible for the misalignments between the 
galaxy stellar and DM spin vectors must develop a different type of connection between the galaxy stellar spins and the local 
tidal fields.  

Given that all baryonic processes gradually and incessantly occur during the nonlinear evolution of galaxies, it would be very difficult to directly identify 
from hydrodynamic simulations the process which originates the peculiar tidal connection of the galaxy stellar spins with the major principal axes of the 
local tidal fields.
Notwithstanding, it may be possible to infer its nature by inspecting the dependence of the galaxy properties on the degree of the misalignments between 
the galaxy stellar and DM spins. If a certain galaxy property is found to be similar to the peculiar tidal connection in their dependence on the misalignment 
degree,  then, the baryonic process responsible for the galaxies to have that particular property must be closely linked with the development of the peculiar 
tidal connection of the galaxy stellar spins. 
Here, we attempt to conduct such an inspection. In section \ref{sec:analy}, we explain which data to use and how to analyze them. 
In section \ref{sec:result}, we describe the results of our analysis and provide physical interpretation of them. In section \ref{sec:con} 
we discuss the implications of the key results and draw a final conclusion.  

\section{Numerical Analysis}\label{sec:analy}

The IllustrisTNG \citep{tngintro1, tngintro2, tngintro3, tngintro4, tngintro5, illustris19} is a suite of the highest-resolution 
cosmological hydrodynamical simulations ever performed by the {\tt AREPO} code \citep{arebo} for the Planck cosmology \citep{planck16}. 
Extensive baryon physics such as radiative cooling, star formation, stellar evolution, chemical enrichment, and feedbacks from supernova 
and growths of blackholes, and so forth were successfully implemented into the IllustrisTNG simulations to keep track of the galaxy evolution. 
The IllustrisTNG 300-1, one of the simulations belonging to the suite, includes in a periodic box of volume $(302.6)^{3}\,$Mpc$^{3}$ 
a total of $2500^{3}$ baryonic particles and equal number of DM particles as well, whose individual masses are $1.1\times 10^{7}\,M_{\odot}$ 
and $5.9\times 10^{7}\,M_{\odot}$, respectively.  The catalogs of the host halos and their subhalos identified at each snapshot of the IllustrisTNG 300-1 
by the standard friends-of-friends (FoF) and subfind algorithms \citep{subfind}, respectively, and their merger trees as well are all available from the 
IllustrisTNG webpage\footnote{https://www.tng-project.org/data/}.
Throughout this paper, we use the two terms, galaxies and subhalos, interchangeably. 

Following the exactly same procedure described in \citet{lee-etal21}, we apply the cloud-in-cell method to the FoF halo catalog from the IllustrisTNG 
300-1 simulation at each of three redshifts, $z=0,0.5$ and $1$, for the construction of the tidal field smoothed on the scale of $R_{f}$, ${\bf T}=(T_{ij})$ 
on $256^{3}$ grids. Given that the majority of the galaxies reside in the group or cluster environments \citep[e.g.,][]{sch-etal97,tem-etal14} the typical scales 
of which are in the range of $1\le R_{f}/{\rm Mpc}\le 2$, the smoothing scale for ${\bf T}$ is set at the median value of the typical cluster sizes, $R_{f}=1.5\,$Mpc. 
At each grid, we calculate three eigenvalues, $\{\lam_{1},\lam_{2},\lam_{3}\}$ with decreasing order, and the corresponding eigenvectors
$\{{\bf e}_{1},{\bf e}_{2},{\bf e}_{3}\}$, which satisfy the condition of ${\bf T}\,{\bf e}_{i}=\lambda_{i}{\bf e}_{i}$ with $i\in \{1,2,3\}$.  
The major principal axis of the tidal field at each grid is in the direction of the eigenvector, ${\bf e}_{1}$, corresponding to the 
largest eigenvalue, $\lam_{1}$. 

While the readers should be referred to \citet{lee-etal21} for a detailed description of the procedure for the determination of  the eigenvector directions of ${\bf T}$, 
it is worth mentioning here that the tidal field constructed from the halo distribution is a good approximation to that from the particle distribution at least on 
the cluster scale. It was indeed found by the previous works that this approximation works especially well in determining the 
directions of the principal axes of the tidal tensors \citep{lee-etal20, LL20}.   Note also that for an application to real observations,  it may be a very reasonable 
choice to adopt this approximation since it is impossible to construct the tidal field from the DM particle distributions in practice \citep{wan-etal09}. 

The IllustrisTNG 300-1 subhalo catalog provides various information on each subhalo at each redshift such as total mass ($\tm$), stellar mass ($\ms$)  
within twice the half-mass radius ($2R_{\star,1/2}$), instantaneous star formation rate (SFR), average metallicity ($Z$) within $2R_{\star,1/2}$, one dimensional 
velocity dispersion ($\sigma_{v}$), central blackhole mass ($M_{\rm bh}$), number of the constituent DM particles ($n_{\rm t}$), number of the constituent stellar 
particles ($n_{\rm s}$) within $2R_{\star,1/2}$, the comoving position of the center (${\bf x}_{c}$) computed as the spatial position of the particle with the minimum 
gravitational potential energy in comoving coordinates, and the peculiar velocity of the center (${\bf v}_{c}$) computed as the sum of the mass weighted velocities of 
all particles in each subhalo.

In addition to these relevant properties of each subhalo, we determine the formation epoch, $a(t_{\rm form})$, 
and stellar shape parameter, $\kappa$. 
For the determination of the former,  we trace the main progenitor of each subhalo by analyzing its merger tree and find the scale factor 
of the epoch when its mass equals $\tm/2$ \citep{merger_tree}. The larger (smaller) the value of $a(t_{\rm form})$ is, the later (earlier) 
a subhalo must have formed. 
For the determination of the latter, we evaluate the fraction of the kinetic energy contributed by its rotational motion within $2R_{\star,1/2}$ for 
each subhalo, as described in \citet{sal-etal12}. The lower (higher) this value of $\kappa$ is relative to the conventional threshold $0.5$, 
the more spheroidal (disk) shape a galaxy must have \citep[see also Equation (7) in ][]{lee-etal21}.

We also extract information on all constituent particles and cells of each subhalo from the snapshot data available at the IllustrisTNG webpage, 
and calculate the DM spin vector, ${\bf J}_{\rm d}$, of each subhalo as
\begin{equation}
\label{eqn:dm_spin}
{\bf J}_{\rm d} = \sum_{i=1}^{n_{\rm m}}{m}_{p,i}\,[({\bf x}_{p,i}-{\bf x}_{c})\times({\bf v}_{p,i}-{\bf v}_{c})]\, , 
\end{equation}
where ${m}_{p,i}$, ${\bf x}_{p,i}$ and ${\bf v}_{p,i}$ are the mass, comoving position and peculiar velocity of the $i$th DM particle that 
constitute each subhalo, respectively.  The unit DM spin vector of each subhalo, $\bjd$, is now obtained as $\bjd\equiv {\bf J}_{d}/\vert{\bf J}_{d}\vert$. 
In a similar manner, we also calculate the unit stellar spin vector, $\bjs$. From here on we refer to $\bjs$ and $\bjd$ as the galaxy stellar 
and DM spin vectors, respectively. 

Recalling the fact that the angular momentum vectors determined from less than $300$ particles are likely to be contaminated by shot noises  
\citep{bet-etal07}, we select only those subhalos with $n_{d}\ge 300$ and $n_{s}\ge 300$ to calculate $\cos\alpha\equiv \vert\bjs\cdot\bjd\vert$.
The smaller the value of $\cos\alpha$ is, the more severely the stellar spin direction of a galaxy deviates from that of its DM counterpart. 
Let $\{\cos\alpha_{1/4}$, $\cos\alpha_{1/2}$, $\cos\alpha_{3/4}$ $\cos\alpha_{1}\}$ denote the values of $\cos\alpha$ whose ranks are  
$\{N_{\rm s}/4$, $N_{\rm s}/2$, $3N_{\rm s}/4$ and $N_{\rm s}\}$, respectively, where $N_{\rm s}$ is the total number of the selected subhalos at each redshfift. 
Then, we make four original samples, $A_{o}$, $B_{o}$, $C_{o}$ and $D_{o}$ of the selected galaxies whose $\cos\alpha$ values lie in the ranges 
of $[0,\cos\alpha_{1/4}]$, $[\cos\alpha_{1/4},\cos\alpha_{1/2}]$, $[\cos\alpha_{1/2},\cos\alpha_{3/4}]$, and $[\cos\alpha_{3/4},\cos\alpha_{1}]$, 
respectively. 

Given that the strengths and types of the alignments between the galaxy spin vectors and the tidal principal directions are known to depend 
on both of the galaxy masses and environmental densities \citep[e.g.,][]{lee-etal20},  it is important and necessary to nullify the effects of the 
mass and density differences among the four samples to find the {\it net} dependence of galaxy properties on $\cos\alpha$. 
Divide the two-dimensional configuration space spanned by $m_{\rm tot}\equiv\log\left[M_{\rm tot}/(10^{7}\,M_{\odot})\right]$ and 
$u\equiv\log\left(1+\delta\right)$ with $\delta\equiv \sum_{i=1}^{3}\lam_{i}$ into tiny pixels of equal area $\Delta\,m_{\rm tot}\Delta\,u$, 
we count the numbers of the galaxies say, $n_{A},\ n_{B},\ n_{C},\ n_{D}$, from the four samples at each pixel. Defining $n_{c}$ as 
$n_{c}\equiv {\rm min}\{n_{A},n_{B},n_{C},n_{D}\}$, we select $n_{c}$ galaxies at each pixel from each of the four original samples to create four new 
samples which are controlled to share identical joint distributions of $m_{\rm tot}$ and $u$. That is, there is no difference among the four samples 
in their total mass and density contrast distributions. From here on, the four controlled samples are referred to as the samples $A, B, C$ and $D$. 
In Table \ref{tab:ngalaxy} are listed the total numbers of the galaxies belonging to the controlled samples ($N_{c}$) as well as 
the mean values of the total mass and linear density contrast at each redshift. 

At the grid where each galaxy from each controlled sample is placed, we compute the cosine of the angle between the galaxy stellar spin vector 
and the major principal axis of the local tidal field, $\cos\theta\equiv\vert\bjs\cdot\be\vert$. Then, we take its 
ensemble average, $\langle\cos\theta\rangle$, separately over each controlled sample at each of the three redshifts.  
The associated errors are calculated as one standard deviation in the determination of $\langle\cos\theta\rangle$ from each sample. 
We also compute the ensemble averages and associated errors of the following galaxy properties over each controlled sample: 
$\langle \ms/\tm\rangle$, $\langle M_{\rm bh}/M_{\star}\rangle$, $\langle {\rm sSFR}\rangle$ with sSFR$\equiv{\rm SFR}/\ms$, 
$\langle a(t_{\rm form})\rangle$, $\langle\kappa\rangle$, $\langle Z\rangle$, and $\langle\sigma_{v}\rangle$. 
These galaxy properties are chosen as the ones being relevant to our analysis since their average values are 
expected to reflect what baryonic processes the galaxies may have undergone during their evolution. 

In a similar manner, we also determine the ensemble average and associated errors of the cosmic web anisotropy parameter, $\xi$, 
defined as \footnote{In the original definition, \citet{ram-etal19} used an adaptive smoothing scale of the tidal field, 
 $R_{f}=4R_{\rm vir}$ where $R_{\rm vir}$ is the virial radius of each halo. But in the current analysis, we use an 
 universal smoothing scale, $R_{f}=1.5\,$Mpc for simplicity.} \citep{ram-etal19}
\beq
\label{eqn:alp}
\xi \equiv \frac{1}{1+\delta}\bigg{\{}\frac{1}{2}\left[(\lam_{1}-\lam_{2})^{2}+
(\lam_{2}-\lam_{3})^{2}+(\lam_{1}-\lam_{3})^{2}\right]\bigg{\}}^{1/2}\, ,
\eeq
over each of the four controlled samples. 
At a fixed region, the value of $\xi$ becomes higher as $z$ decreases since it reflects how nonlinearly the tidal field has 
evolved.  At a fixed redshift, relatively higher values of $\xi$ are expected from the regions where the tidal fields 
are coherent over large scales, while relatively lower values from those where the tidal fields do not possess 
a large-scale coherence \citep[e.g.,][]{lee19,zju-etal20}. For example, at a given low redshift ($z<1$) when the complex filamentary 
web-like structures emerge through nonlinear evolution of the tidal fields \citep{web96},  the intersections of multiple 
narrow secondary filaments would yield lower values of $\xi$, while the regions inside single thick primary filaments would 
yield higher values of $\xi$ \citep{zomg1}. 
 
\section{Results and Interpretations}\label{sec:result}

\begin{figure}[ht]
\begin{center}
\plotone{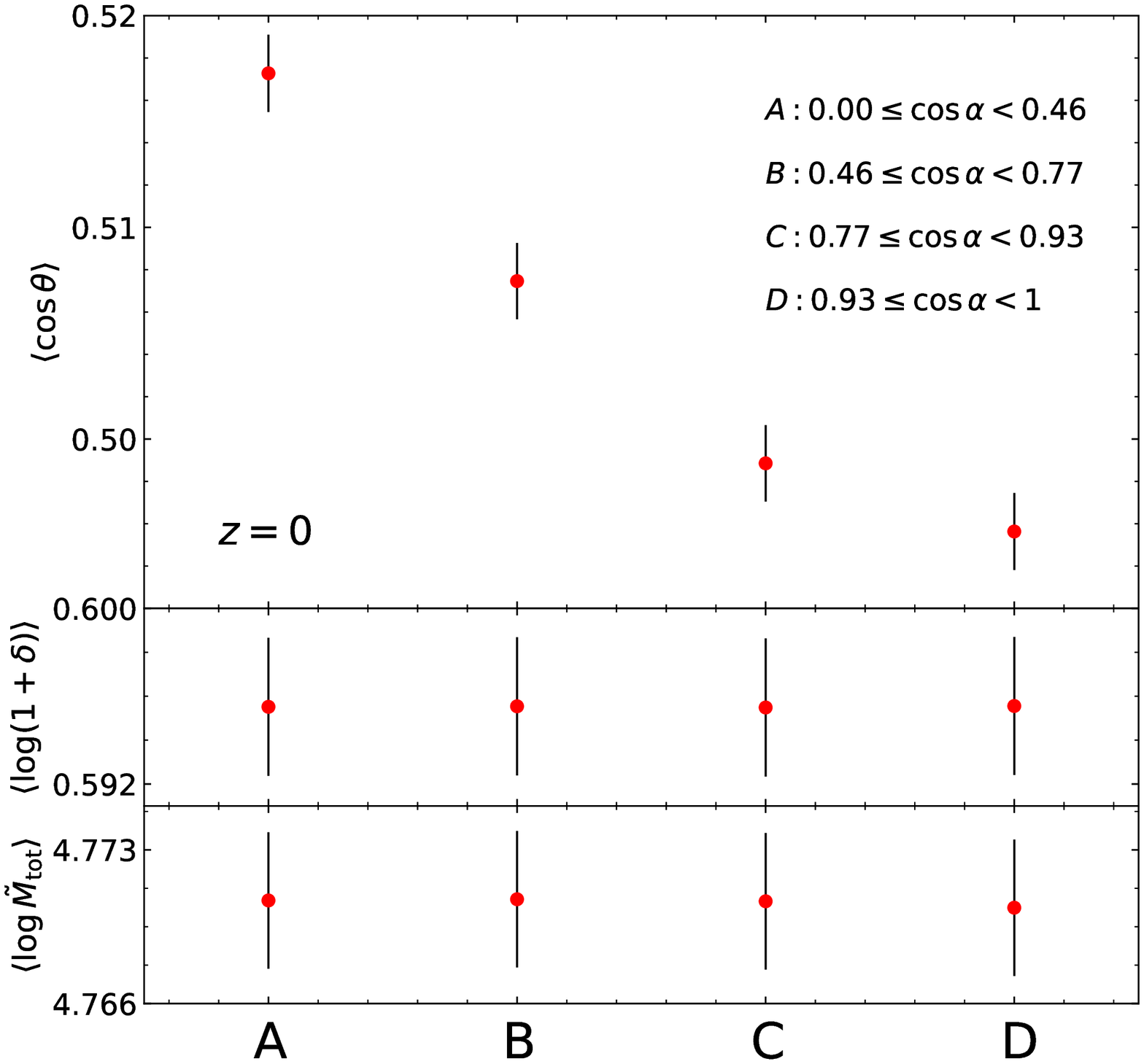}
\caption{Mean cosines of the angles between the galaxy stellar spins and the major principal axes of the local tidal fields at $z=0$, 
from four controlled samples (top panel), which have distinct distributions of $\cos\alpha$ which denote the cosines of the angles 
between the galaxy stellar and DM spins. Mean values of $\log (1+\delta)$ and 
$\log\left[M_{\rm tot}/(10^{7}M_{\odot})\right]$, of the four controlled samples (middle and bottom panels, respectively)}
\label{fig:cost_z0}
\end{center}
\end{figure}

Figure \ref{fig:cost_z0} plots $\costa$ (top panel), $\langle m_{\rm tot}\rangle$  (middle panel) and 
$\langle\log(1+\delta)\rangle$ (bottom panel) from the samples, $A$, $B$, $C$ and $D$, at $z=0$.  
As can be seen,  in spite of no difference in $\langle m_{\rm tot}\rangle$ and $\langle\log(1+\delta)\rangle$ among the four 
$\cosa$-selected samples, a clear signal of the existence of a monotonic anti-correlation between $\costa$ and $\cosaa$ is detected. 
The values of $\costa$ from the samples $A$ and $B$ in the low-$\cosa$ range are significantly higher than $0.5$, indicating that the 
stellar spin vectors from these two samples are preferentially aligned with the major principal directions of the local tidal fields. 
The more severely the stellar spin vectors of the galaxies deviate from those of the DM counterparts, the more strongly they 
tend to be aligned with the major principal directions of the local tidal fields.
This $\bjs$-$\be$ alignment tendency found from the samples $A$ and $B$ is peculiar in the respect that the halo DM or total spin vectors 
exhibit a strong tendency of being aligned with either the intermediate or the minor principal axes of the tidal fields but being persistently 
perpendicular to the major principal axes in the entire mass range at all redshifts as shown in \citet{lee-etal21}. 

Meanwhile, the values of $\costa$ from the samples $C$ and $D$ show no significant difference from $0.5$, indicating that the stellar spin 
orientations of the galaxies from these two samples in the high $\cosa$ range are random relative to the tidal principal directions. 
This result implies that there may be some threshold on the value of $\cosa$ for the existence of the peculiar $\bjs$-$\be$ alignments. 
In other words, only if the $\bjs$-$\bjd$ alignments drops below some threshold value, the peculiar $\bjs$-$\be$ alignments can be developed 
at $z=0$. 

\begin{figure}[ht]
\begin{center}
\plotone{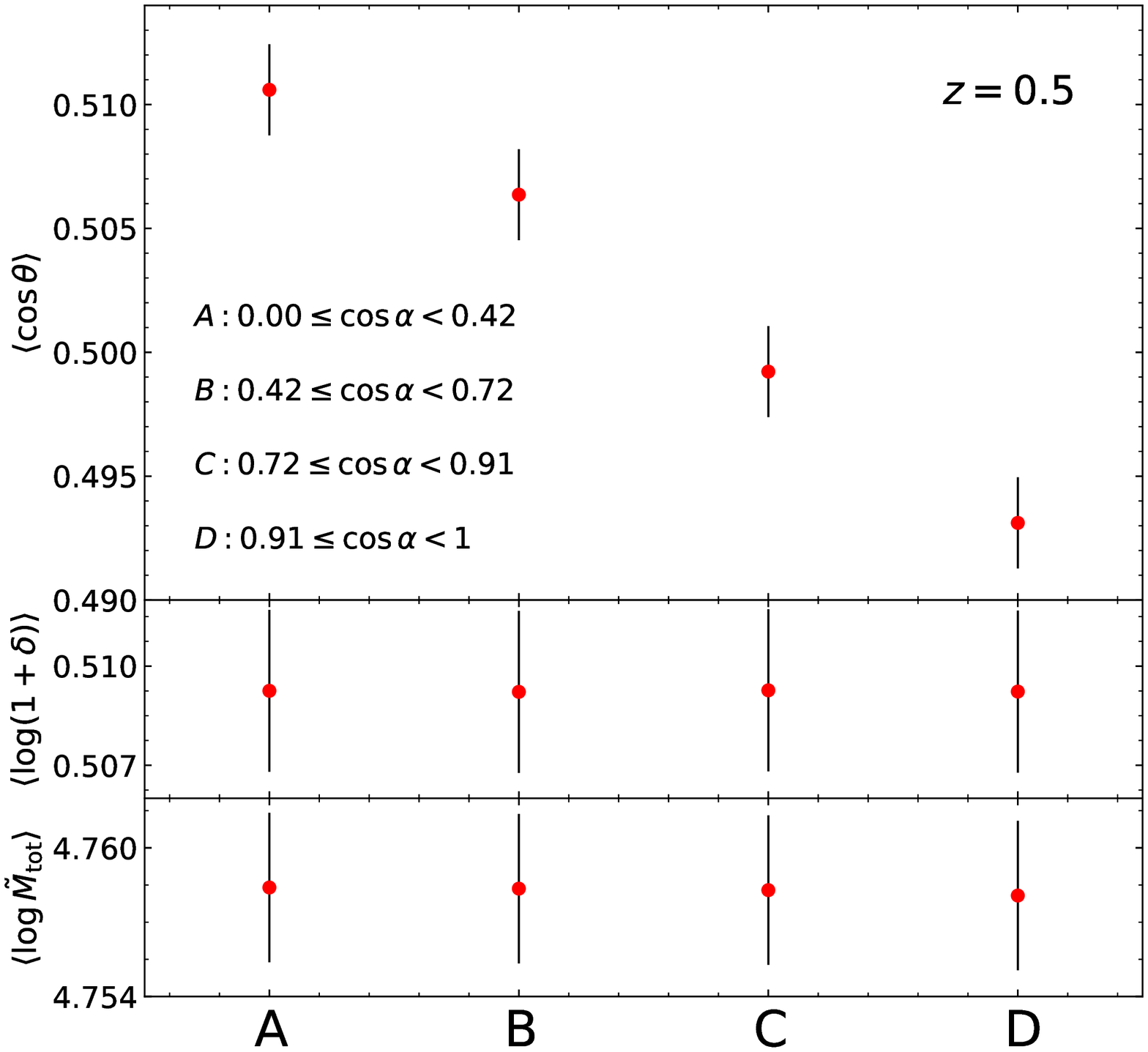}
\caption{Same as Figure \ref{fig:cost_z0} but at $z=0.5$.}
\label{fig:cost_z0.5}
\end{center}
\end{figure}

\begin{figure}[ht]
\begin{center}
\plotone{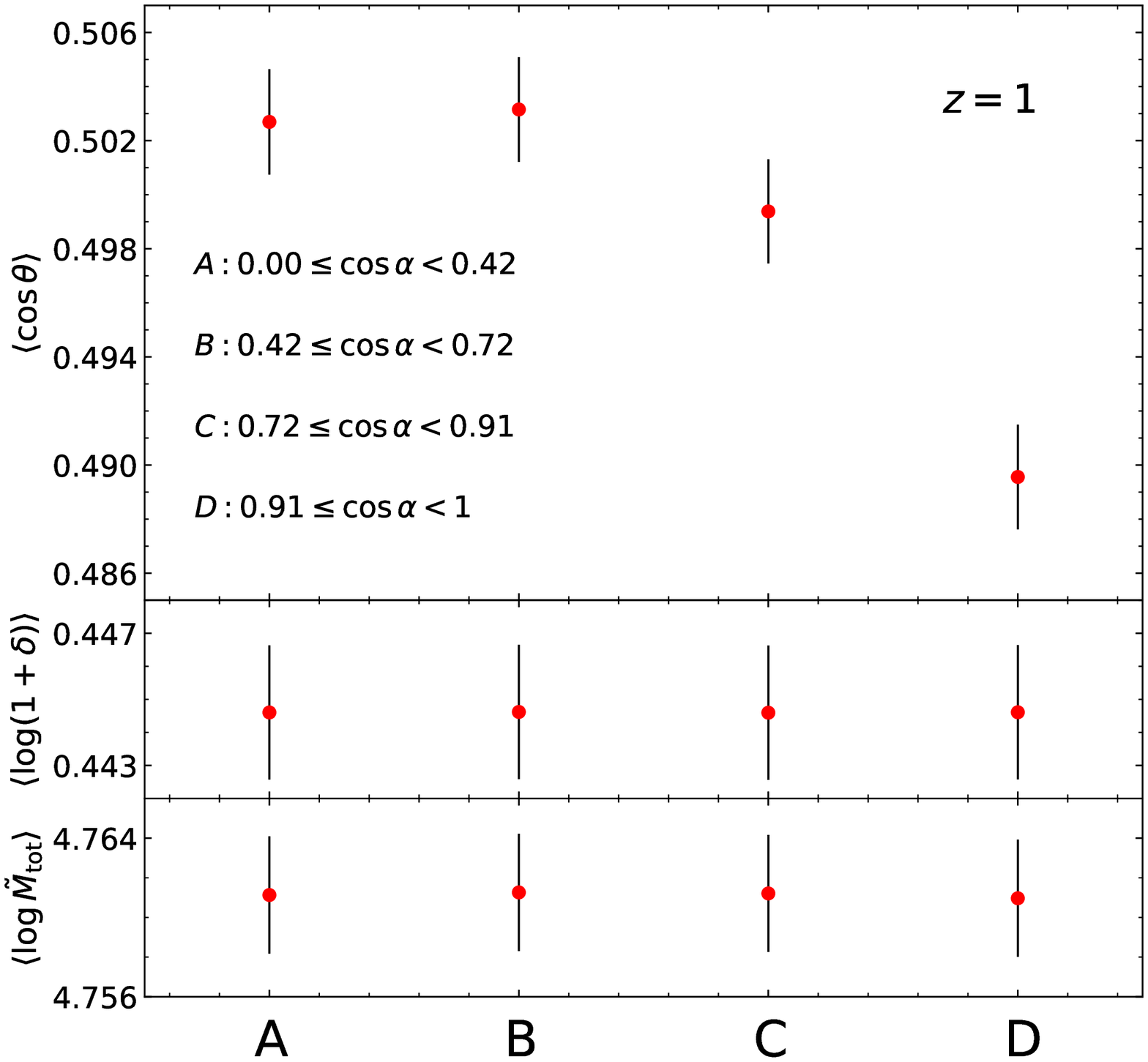}
\caption{Same as Figure \ref{fig:cost_z0} but at $z=1$.}
\label{fig:cost_z1}
\end{center}
\end{figure}

Figures \ref{fig:cost_z0.5} and \ref{fig:cost_z1} plot the same as Figure \ref{fig:cost_z0} but at $z=0.5$ and $1$, respectively. 
The results at $z=0.5$ are similar to those at $z=0$ in the respect that a monotonic anti-correlation between $\costa$ and $\cosaa$ is found 
from the samples $A$ and $B$. But, the peculiar $\bjs$-$\be$ alignments from those two samples at $z=0.5$ are much weaker than at $z=0$. 
Moreover, the $\costa$ value from the sample $D$ at $z=0.5$ are substantially lower than $0.5$, signalling a weak tendency of being perpendicular 
to the tidal major principal axes.
At $z=1$, a signal of the peculiar $\bjs$-$\be$ alignment tendency is found from none of the four samples. The stellar spin vectors of the galaxies 
from the samples $A$, $B$ and $C$ seem to be random relative to the tidal principal axes, while those from the sample $D$ show 
a strong tendency of being perpendicular to the tidal major principal axes. Similar results are also found at $z=1.5$ and $z=2$ but with larger errors 
due to the small sizes of the controlled samples at these higher redshifts. 
Give the results shown in Figure \ref{fig:cost_z0}-\ref{fig:cost_z1}, we suggest that the main baryonic processes responsible for the $\bjs$-$\bjd$ misalignments 
should occur somehow differently between the low and high redshifts.  At $z\ge 1$, its effect is only diminishing any tidally induced alignments of $\bjs$. 
Whereas, at $z\le 0.5$, it has an effect of causing the peculiar alignment of $\bjs$ with $\be$. 

\begin{figure}[ht]
\begin{center}
\plotone{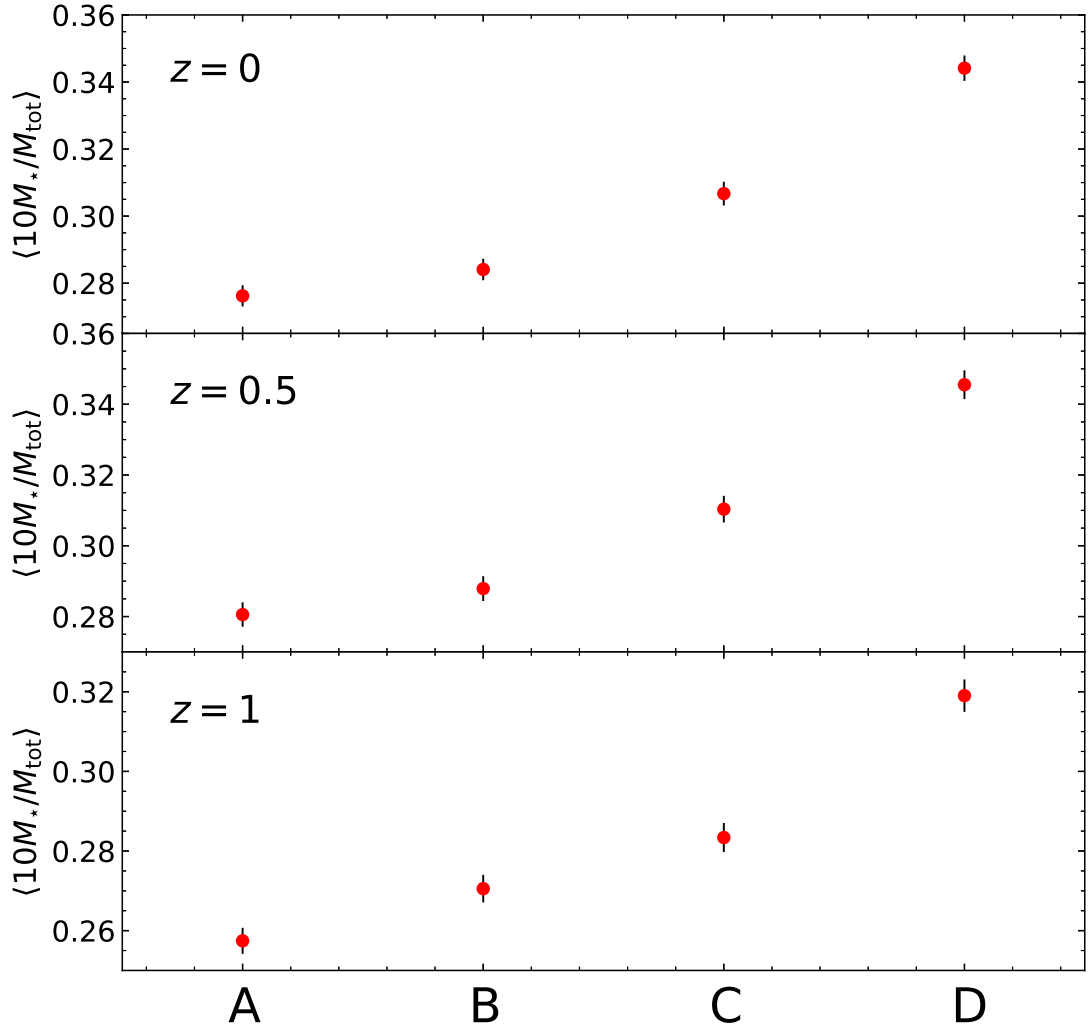}
\caption{Ensemble averages of the stellar-to-DM mass ratios taken over each of the four controlled samples 
at three different redshifts.}
\label{fig:msratio}
\end{center}
\end{figure}

\begin{figure}[ht]
\begin{center}
\plotone{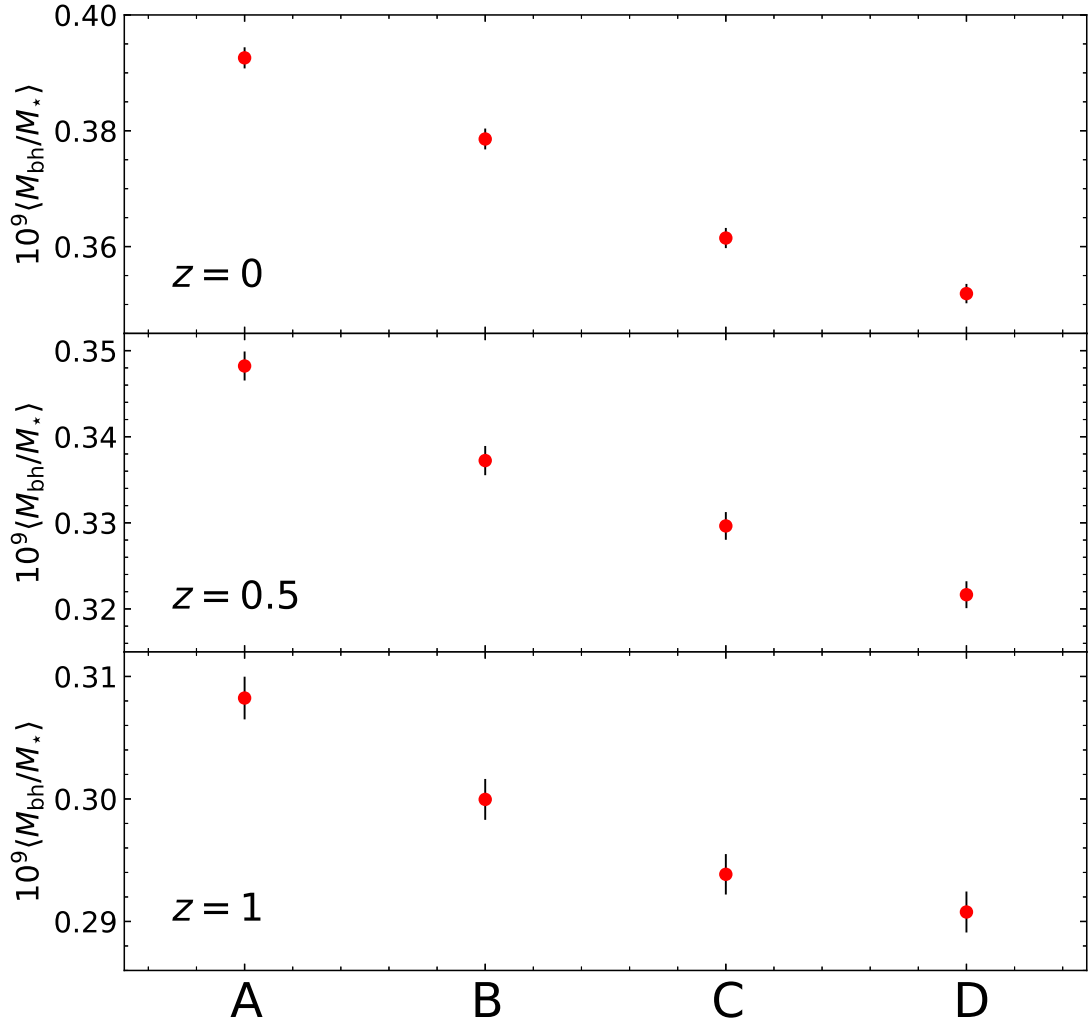}
\caption{Ensemble averages of the blackhole-to-stellar mass ratios taken over each of the four controlled samples 
at $z=0, 0.5$ and $1$.}
\label{fig:mbh}
\end{center}
\end{figure}

\begin{figure}[ht]
\begin{center}
\plotone{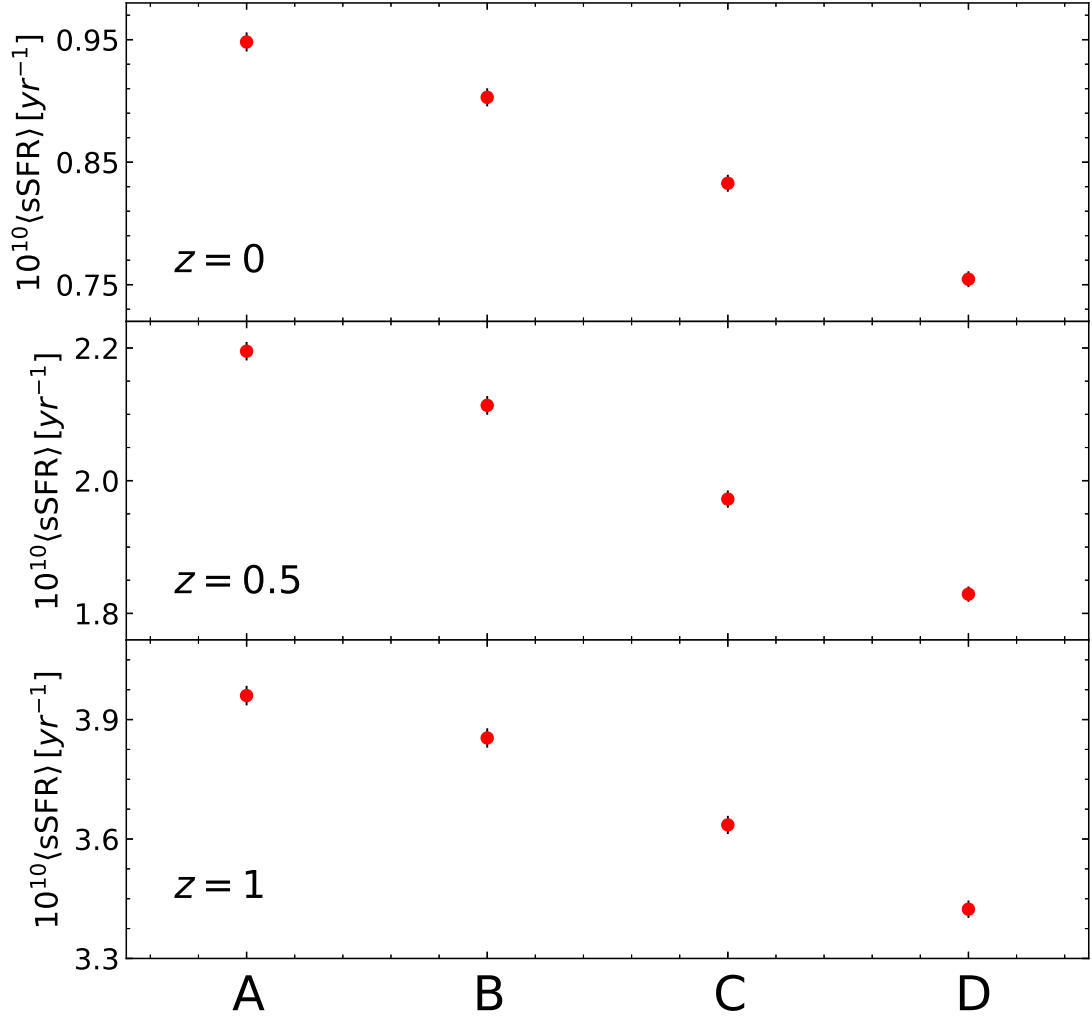}
\caption{Ensemble averages of the specific star formation rates taken over each of the four controlled samples 
at $z=0, 0.5$ and $1$.}
\label{fig:sfr}
\end{center}
\end{figure}

\begin{figure}[ht]
\begin{center}
\plotone{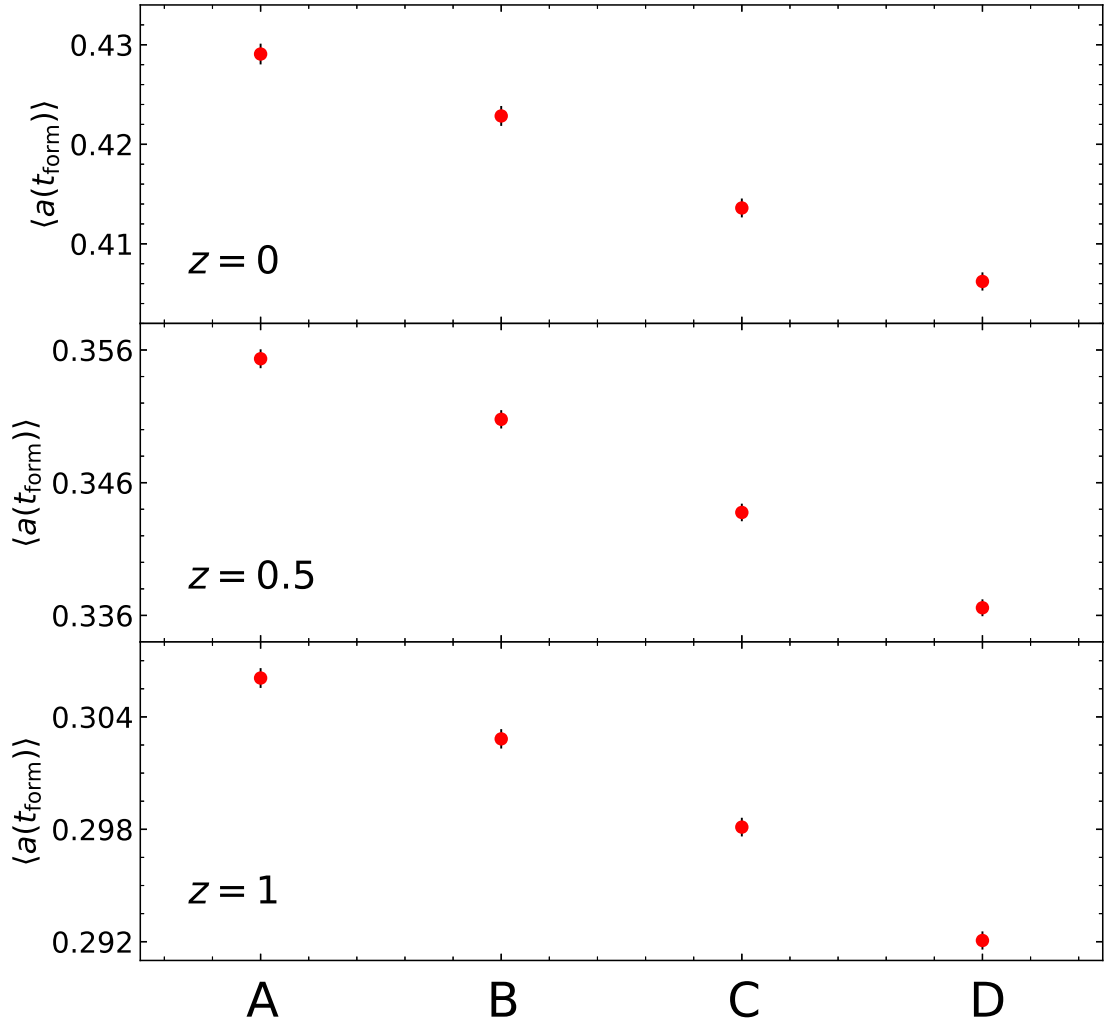}
\caption{Ensemble averages of the formation epochs taken over each of the four controlled samples 
at $z=0, 0.5$ and $1$.}
\label{fig:form}
\end{center}
\end{figure}

\begin{figure}[ht]
\begin{center}
\plotone{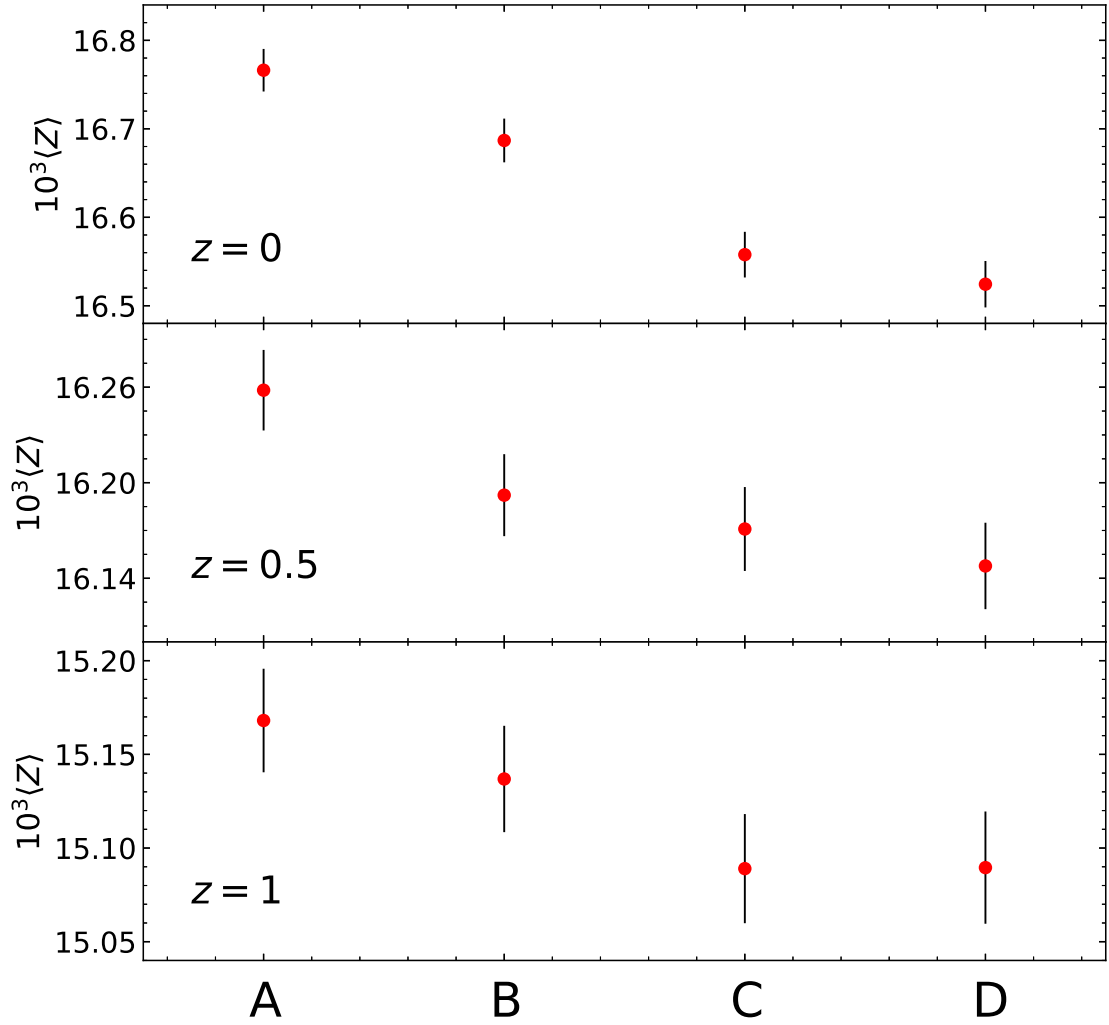}
\caption{Ensemble averages of the metallicities taken over each of the four controlled samples 
at $z=0, 0.5$ and $1$.}
\label{fig:metal}
\end{center}
\end{figure}

\begin{figure}[ht]
\begin{center}
\plotone{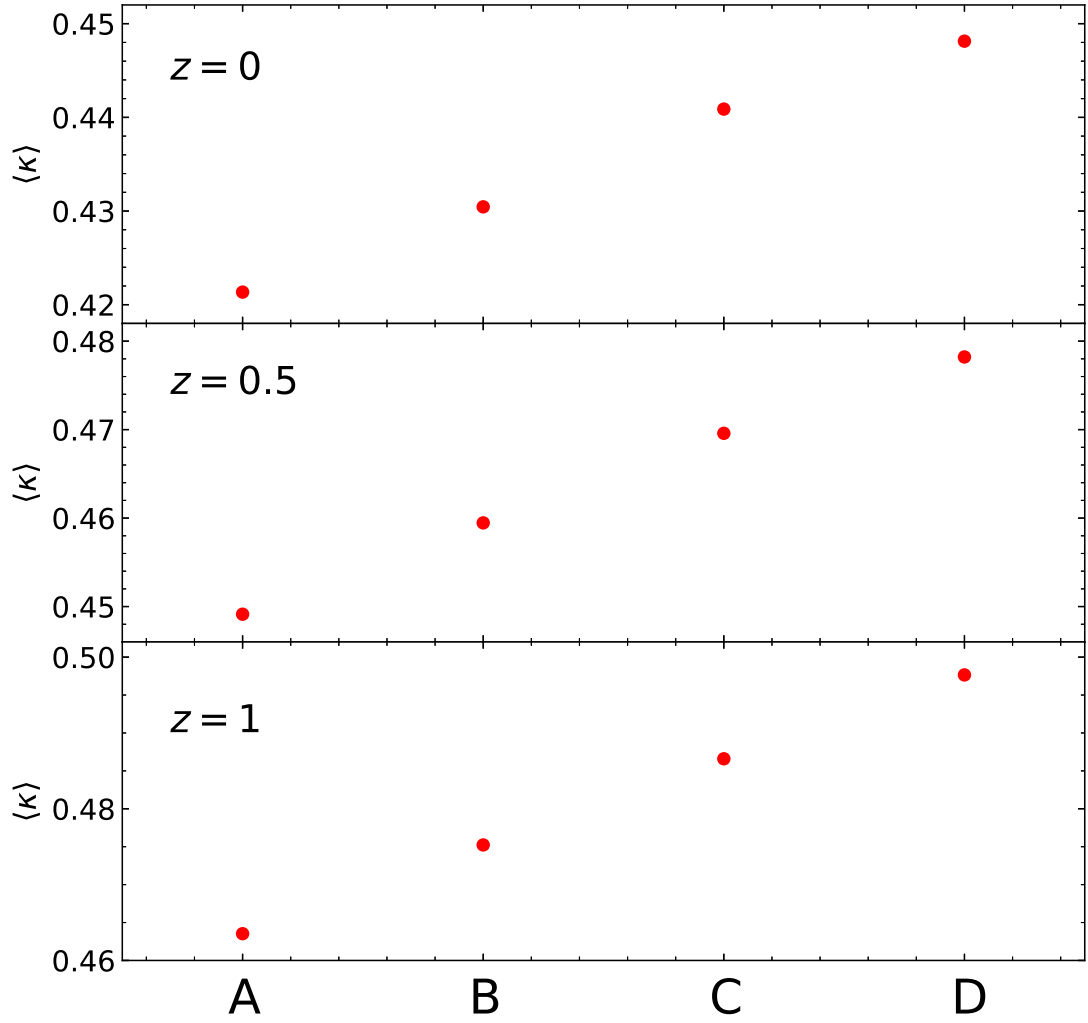}
\caption{Ensemble averages of the fraction of kinetic energy contributed by the rotation motions taken over each of the four controlled samples 
at $z=0, 0.5$ and $1$.}
\label{fig:kappa}
\end{center}
\end{figure}

\begin{figure}[ht]
\begin{center}
\plotone{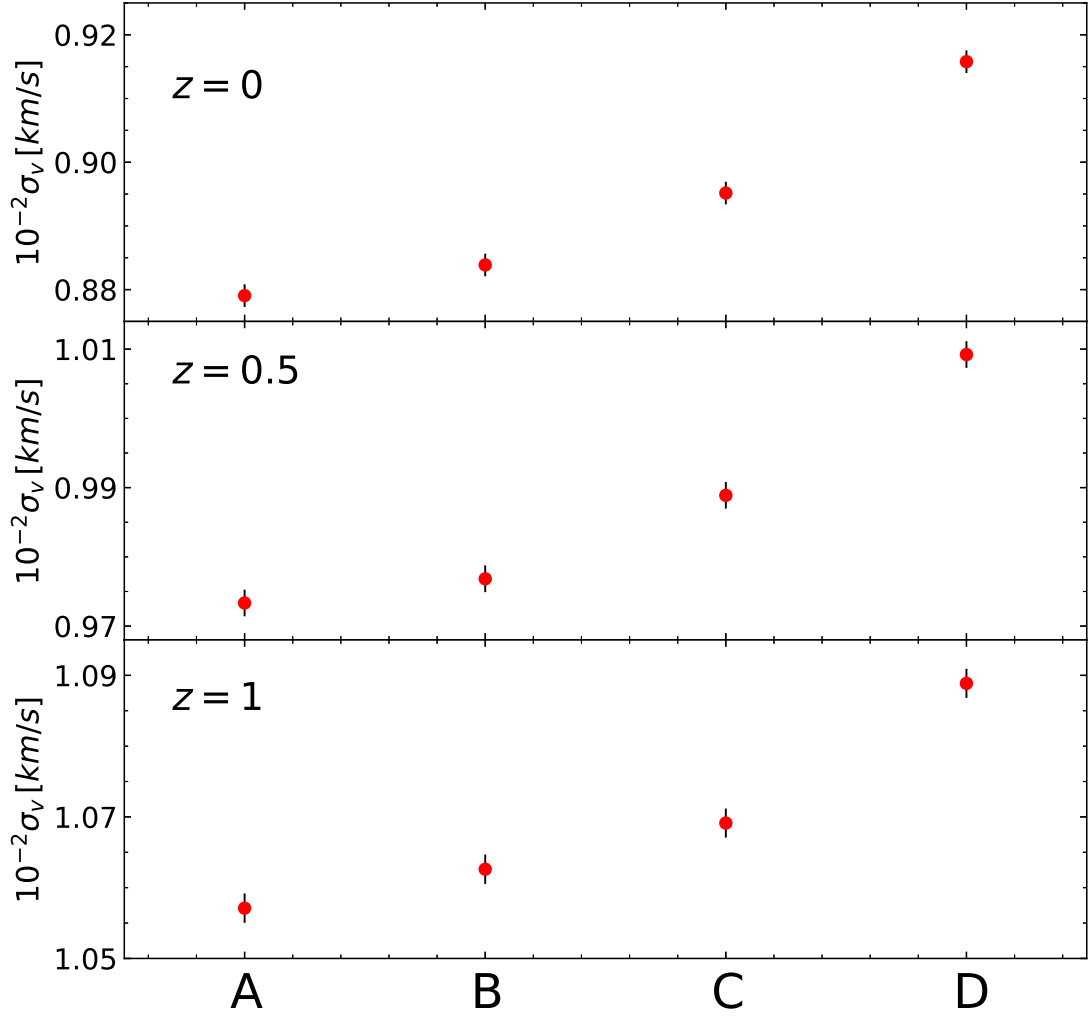}
\caption{Ensemble averages of the velocity dispersions taken over each of the four controlled samples 
at $z=0, 0.5$ and $1$.}
\label{fig:vdis}
\end{center}
\end{figure}

\begin{figure}[ht]
\begin{center}
\plotone{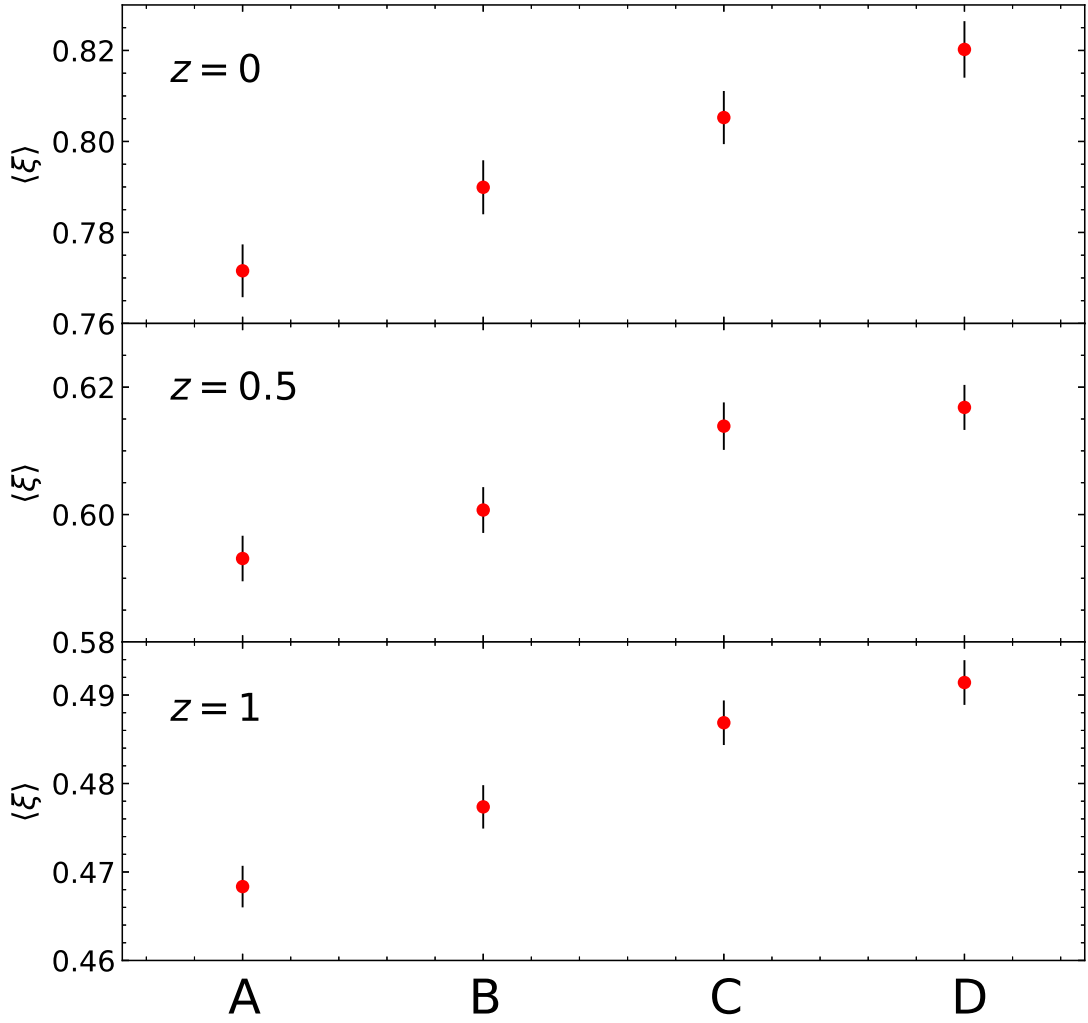}
\caption{Ensemble averages of the cosmic web anisotropy parameter values taken over each of the four controlled samples 
at $z=0, 0.5$ and $1$.}
\label{fig:web}
\end{center}
\end{figure}

Figures \ref{fig:msratio}-\ref{fig:web} plot $\langle \ms/\tm\rangle$,$\langle M_{\rm bh}/\ms\rangle$,$\langle{\rm sSFR}\rangle$, $\langle a(t_{\rm form})\rangle$, 
$\langle Z\rangle$, $\langle \kappa\rangle$, $\langle\sigma_{v}\rangle$ and $\langle\xi\rangle$, from the four controlled samples at the three redshifts, respectively.
As can be seen, the existence of strong anti-correlation with $\cosa$ is exhibited by $M_{\rm bh}/\ms$,${\rm sSFR}$, $a(t_{\rm form})$,  while strong correlation
with $\cosa$ by $\ms/\tm$, $\kappa$\, $\sigma_{v}$ and $\xi$ at all of the three redshifts. As for the metallicity $Z$, it shows strong anti-correlation with $\cosa$ 
at $z=0$, weak anti-correlation at $z=0.5$, but no significant signal of anti-correlation nor correlation at $z=1$, which behavior is similar to that of $\cost$. 

The results shown in Figures \ref{fig:msratio}-\ref{fig:web} imply the following. Whatever baryonic process originates the misalignments between the 
galaxy stellar and DM spin directions,  it must be most effective in the relatively less anisotropic web environments. The galaxies located in these lower 
$\xi$-regions can more efficiently accrete materials \citep[e.g.,][]{zomg1}, which lead them to have more massive central blackholes, later formation 
epochs, higher specific star formation rates, and higher metallicities. They must be also more vulnerable to the feedbacks from more massive blackholes, 
the galactic winds from which would discharge stellar materials from their outer layers \citep[][and references therein]{ten-etal17}, causing them to have lower 
stellar-to-total mass ratios, more spheroidal shapes and lower velocity dispersions. 
The galaxies located in the relatively lower $\xi$-regions at $z\le 0.5$ must also have on average higher metallicities since they tend to form later, accreting gas-richer 
materials unlike the ones located in the relatively higher $\xi$-regions, form earlier where the gas materials were rather metal poor. No detection of anti-correlation signal 
between $\langle Z\rangle$ and $\cosaa$ at $z=1$ can be understood by the same logic. 
At that high redshift, all material accreted by the galaxies are metal poor, no matter what their formation epochs are.

Note the redshift variation of $\xi$ in Figure \ref{fig:web}. The overall values of the cosmic anisotropy parameter are much lower at higher redshifts, 
revealing that the cosmic web becomes more and more anisotropic through the nonlinear evolution of the tidal fields.  Recalling that the peculiar 
$\bjs$-$\be$ alignments can be found only at $z\le 0.5$ but not at $z=1$ and that its strength sharply increases as $z$ decreases from $z=0.5$ to $0$ 
(Figures \ref{fig:cost_z0}-\ref{fig:cost_z1}),  we interpret the results shown in Figure \ref{fig:web} as an evidence for the {\it anisotropic} occurrence 
of the baryonic process responsible for the $\bjs$-$\bjd$ misalignments at low redshifts. 
At higher redshifts where the overall cosmic web anisotropy is lower, the baryonic process like the galactic winds would occur isotropically, and thus no 
preferential directions  exist for the discharge of stellar materials. 
At lower redshifts where the overall cosmic web has significantly higher anisotropy than a certain threshold, the galactic winds would blow anisotropically along 
the major principal axes of the local tidal fields, discharging stellar materials more severely along that $\be$-directions, generating the peculiar $\bjs$-$\be$ 
alignments.  In short, the development of the peculiar $\bjs$-$\be$ alignments would be subject to the anisotropic occurrence of the the stronger baryonic 
feedbacks. Although the stronger baryonic feedbacks would occur preferentially in the lower $\xi$ regions at a fixed redshift, its anisotropic occurrence would be 
possible only at lower redshifts when the degree of the cosmic web anisotropy exceeds some threshold. 

It is worth mentioning here that our results are robust against the variation of the smoothing scale in the range of $1.5\le R_{f}/{\rm Mpc}\le 2$. 
As already confirmed by \citet{lee-etal21},  the change of $R_{f}$ does not alter the tendency of the tidally induced alignments of the galaxy stellar spins, 
although the alignment strengths tend to decrease as $R_{f}$ increases. 
It is also worth mentioning that the samples used for our analysis contain not only the galaxy size subhalos but also the group and cluster size subhalos 
with stellar masses larger than $5\times 10^{12}\,\munit$.  Our results, however, have been shown to be robust, regardless of their existence. 
Even when we eliminate those massive subhalos, we end up having the essentially same results, albeit with slightly larger errors. 

\section{Summary and Discussion}\label{sec:con}

It has been conventionally thought that all nonlinear baryonic processes must have an effect of erasing or at least weakening the retaining link between any 
galaxy properties and the initial conditions of the universe. That was why the prospect of using the tidally induced spin alignments of galactic halos  
as a probe of cosmology was thought to be largely contingent upon how strong effects the baryonic processes have on the galaxy spin orientations. 
If the effects are strong enough to significantly deviate the galaxy stellar spin directions from those of the DM counterparts, then the {\it observable} 
former would become insensitive to the initial conditions due to the loss of tidal connections that the {\it unobservable} latter retains.

This conventional thought, however, has been challenged by the very recent finding of \citet{lee-etal21} that the stellar spins of massive 
galaxies at low redshifts exhibit a peculiar tendency of being aligned with the major principal axes of the local tidal fields (dubbed, a peculiar 
tidal connection), in contrast to their DM counterparts that always exhibit a strong tendency of being orthogonal to the tidal major axes in the entire 
mass section.  
To understand how any baryonic process could originate this peculiar tidal connection of the galaxy stellar spins, we have numerically 
investigated the {\it net} dependences of various properties of the galaxies from the IllustrisTNG 300-1 hydrodynamic simulations on the 
degree of the alignments, $\cos\alpha$, between the orientations of galaxy stellar and DM spins at three different redshifts, $z=0,0.5$ and $1$. 
The possible effects of the mass and density differences have been all controlled down to a negligible level for this investigation, the results of 
can be summarized in the following.
\begin{itemize}
\item
At $z\le 0.5$ the galaxies whose stellar spin directions deviate more severely from those of their DM spins (i.e., the galaxies with lower mean values of 
$\cos\alpha$) yield higher degree of the spin alignments with the major principal axes of the local tidal fields. 
The strength of the $\cos\alpha$-dependence of this peculiar tidal connection is found to decrease as $z$ increases.  
At $z=1$, the galaxy stellar spins have random orientations with respect to the tidal principal axes in the low-$\cosa$ range while they show a weak signal of being 
perpendicular to the tidal major axes, similar to the DM counterparts in the highest-$\cosa$ range. 
\item
At all of the three redshifts, the galaxies with lower $\cos\alpha$  have on average the following properties: 
lower stellar-to-DM mass ratios, higher blackhole-to-stellar mass ratios, higher specific star formation rates, later formation epochs, 
more spheroidal shapes, lower velocity dispersions, and being located in the regions with lower degree of the cosmic web anisotropy.
The strengths of the $\cos\alpha$-dependence of these properties are robust against the redshift-variation.
\item 
The $\cosa$-dependence of the galaxy metallicity exhibits the most similar redshift variation to that of the peculiar tidal connections. 
At $z\le 0.5$, the galaxies with lower $\cos\alpha$ have on average significantly higher metallicities. At $z=1$, 
however, no significant signal of the $\cos\alpha$-dependence of the galaxy metallicity is found. 
\end{itemize}

It is worth mentioning here that although the detected signal is statistically significant, the peculiar tidal connection of the galaxy stellar 
spins is quite a weak tendency which would be detectable only from very large galaxy samples. That is probably why it had not been found by the
previous works, which utilized smaller box-size hydrodynamic simulations than the IllustrisTNG 300-1 \citep[e.g.,][]{wan-etal18}.  There are, however, several factors 
other than the sample size, which can affect the signal strength. One might expect a stronger signal of the peculiar tidal connection of the galaxy stellar 
spins if the tidal field were nonlinearly constructed and smoothed on a much smaller scale \citep{lee-etal21}, which 
requires a hydrodynamic simulation with much higher mass resolution and thus beyond the scope of this work. 

To physically explain our results, we put forth the following scenario.
For the galaxies located in the cosmic web environments with relatively lower degree of anisotropy at a fixed redshift, the incessant {\it radial} infall  
of gas and matter particles would have an effect of delaying their formation epochs and enhancing the growths of their central blackholes, 
specific star formation rates and average metallicities \citep{zomg1}. On the other hand, the galaxies in such environments should be 
more vulnerable to the stronger effects of blackhole feedbacks in in the form of galactic winds, which can take away substantial amounts of stellar materials from  
the outer layers. In consequence, they would have on average lower stellar-to-DM ratios, more spheroidal shapes, and lower velocity dispersions, 
with their stellar spin directions more severely deviating from those  of their DM counterparts. 

At higher redshifts ($z\ge 1$) where the tidal field is still in the quasi-linear stage and the overall cosmic web anisotropy is quite low,  
the baryonic feedbacks would occur quite isotropically in the regions with relatively lower anisotropy.  In such environments, the stellar 
spin directions of the galaxies affected by the stronger baryonic feedbacks would just become random, 
losing the tidally induced alignment tendency that their DM counterparts retain. 
The overall amplitude of the cosmic web, however, gradually increases as the universe evolves. At lower redshifts ($z\le 0.5$) where the 
tidal field has undergone nonlinear evolution, even those web environments with relatively lower anisotropy can develop quite filamentary 
features like the intersections of multiple narrow filaments)  In such environments at $z\le 0.5$, the baryonic feedbacks responsible 
for the discharge of stellar materials from the galaxies may occur preferentially along the major principal directions 
of the tidal fields.  In consequence, the stellar spin directions of the galaxies do not only deviate from those of the DM counterparts but also 
acquire a peculiar tendency of being aligned with the tidal major axes.

The bottom line of our scenario is that the peculiar alignments of the galaxy stellar spins with the tidal major axes at $z\le 0.5$ should be developed by the 
anisotropic occurrence of nonlinear baryonic feedbacks responsible for the discharge of stellar angular momentum along the tidal major axes. 
Although no direct evidence has been presented here, our detection of the $\cosa$-dependences of the galaxy properties and its redshift variation can be 
well explained and coherently understood in this scenario. 
The peculiar tidal alignments might mediate between the galaxy stellar spins and the initial conditions of the universe, even though 
they are acquired through deeply nonlinear baryonic process. 
Our future work will be in the direction of finding a direct evidence for this scenario, statistically modeling the anisotropic occurrence of the baryonic 
feedbacks in the nonlinear cosmic web, and exploring its link to the initial conditions of the universe. 

\acknowledgments

The IllustrisTNG simulations were undertaken with compute time awarded by the Gauss Centre for Supercomputing (GCS) under 
GCS Large-Scale Projects GCS-ILLU and GCS-DWAR on the GCS share of the supercomputer Hazel Hen at the High Performance 
Computing Center Stuttgart (HLRS), as well as on the machines of the Max Planck Computing and Data Facility (MPCDF) in Garching, 
Germany.
We thank our referee for helpful comments and suggestions. 
J.L. acknowledges the support by Basic Science Research Program through the National Research Foundation (NRF) of Korea 
funded by the Ministry of Education (No.2019R1A2C1083855). J.-S. M. acknowledges support from the Basic Science Research 
Program (No. 2021R1A6A3A01086838) through the NRF of Korea funded by the Ministry of Education. 
S.-J. Y. acknowledges support from the Mid-career Researcher Program (No. 2019R1A2C3006242) through the NRF of Korea.

\clearpage
\clearpage
\clearpage
\clearpage
\clearpage
\clearpage
\clearpage
\clearpage
\clearpage
\clearpage
\clearpage
\clearpage
\begin{deluxetable}{cccc}
\tablewidth{0pt}
\setlength{\tabcolsep}{3mm}
\tablecaption{Total Numbers and Mean Density and Mass of the Selected Subhalos}
\tablehead{redshifts & $N_{c}$ & $\log (1+\delta)$ & $\log[M_{\rm tot}/(10^{7}M_{\odot})]$}
\startdata
$0$ & $102496$ & $0.595\pm 0.003$  & $4.770\pm 0.003$ \\
$0.5$ & $122380$ & $0.509\pm 0.002$ & $4.758\pm 0.003$ \\
$1$ & $126328$ & $0.445\pm 0.002$ & $4.761\pm 0.003$\\
\enddata
\label{tab:ngalaxy}
\end{deluxetable}

\end{document}